\documentclass[twocolumn,aps,showpacs,prb,tightenlines,amsmath,amssymb]{revtex4}
\usepackage{graphicx}
\usepackage{amssymb}
\usepackage{dcolumn}
\usepackage{amsmath}
\usepackage{bm}
\usepackage{colordvi}
\newcommand{\bgreek}[1]{\mbox{\boldmath$#1$\unboldmath}}

\begin{document}

\title{Hole spin relaxation in $p$-type (111) GaAs quantum wells}
\author{L. Wang}
\affiliation{Hefei National Laboratory for Physical Sciences at
Microscale and Department of Physics,
University of Science and Technology of China, Hefei,
Anhui, 230026, China}
\author{M. W. Wu}
\thanks{Author to whom correspondence should be addressed}
\email{mwwu@ustc.edu.cn.}
\affiliation{Hefei National Laboratory for Physical Sciences at
Microscale and Department of Physics, University of Science and 
Technology of China, Hefei, Anhui, 230026, China}

\date{\today}

\begin{abstract}
Hole spin relaxation in $p$-type (111) GaAs quantum wells is 
investigated in the case with only the
lowest hole subband, which is heavy-hole like in (111) GaAs/AlAs and
light-hole like in (111) GaAs/InP quantum wells, being relevant. The subband
L\"{o}wdin perturbation method is applied to obtain the effective Hamiltonian
including the Dresselhaus and Rashba spin-orbit couplings. Under a proper gate
voltage, the total in-plane effective magnetic
field in (111) GaAs/AlAs quantum wells can be strongly suppressed in the whole
momentum space, while the one in (111) GaAs/InP quantum wells can be suppressed only on a
special momentum circle. The hole spin relaxation due to the
D'yakonov-Perel' and Elliott-Yafet mechanisms is calculated by means of the fully microscopic
kinetic spin Bloch equation approach with all the relevant scatterings
explicitly included. For (111) GaAs/AlAs quantum wells, extremely long
heavy-hole spin relaxation time (upto hundreds of nanoseconds) is predicted. In
addition, we predict a pronounced peak in the gate-voltage
dependence of the heavy-hole spin relaxation time due to the D'yakonov-Perel'
mechanism. This peak origins from the suppression of the unique
inhomogeneous broadening in (111) GaAs/AlAs quantum wells. Moreover, the Elliott-Yafet
mechanism influences the spin relaxation only around the peak area due to the small spin
mixing between the heavy and light holes in quantum wells with small
well width. We also show the anisotropy of
the spin relaxation. In (111) GaAs/InP quantum wells, 
a mild peak, similar to the case for electrons in
  (111) GaAs quantum wells,  is also predicted in the gate-voltage
dependence of the light-hole spin relaxation time. 
The contribution of the Elliott-Yafet mechanism 
is always negligible in this case. 
\end{abstract}


\maketitle
\section{INTRODUCTION}
Spin dynamics in semiconductor quantum wells (QWs) has been intensively investigated 
in recent years due to the fast development of semiconductor 
spintronics.\cite{meier,wolf,semi,fab,wu,korn,handbook}
In III-V zinc-blende QW systems, spin relaxation is mainly determined by the 
D'yakonov-Perel' (DP) mechanism,\cite{dp} 
which results from the momentum scattering together with the momentum-dependent 
effective magnetic field induced by the Dresselhaus\cite{dresselhaus} and/or Rashba\cite{rashba}
spin-orbit couplings (SOCs). So far, most studies focus on the 
$(001)$\cite{averkiev1,wmw,mali,averkiev2,schliemann,weng,kainz,schliemann2,
winkler1,erlingsson,averkiev3,cheng,zhou,mohno,liu,stich,koralek,yzhou2} 
and $(110)$ QWs.\cite{ohno,lau2,wuku,karimov,hall,dohrmann,yzhou} In these works, many efforts
have been devoted to obtain a long spin relaxation time (SRT).\cite{averkiev1,averkiev2,schliemann,
schliemann2,winkler1,erlingsson,averkiev3,cheng,mohno,liu,stich,
koralek,ohno,hall,dohrmann} In $(001)$ QWs, when the strengths of the Dresselhaus 
and Rashba terms are comparable, electron spin relaxation along the $[110]$ or 
$[1\bar{1}0]$ direction 
can be strongly suppressed.\cite{averkiev1,averkiev2,schliemann,schliemann2,winkler1,
erlingsson,averkiev3,cheng,mohno,liu,stich,koralek}
For $(110)$ symmetric QWs, the effective magnetic field due to the Dresselhaus term 
is oriented along the growth direction, which leads to an absent DP relaxation 
for electron spins along this direction.\cite{ohno,hall,dohrmann}

Recently, some attention has been devoted to $(111)$ QWs where electron spin relaxation 
also shows rich properties.\cite{cartoixa,vurgaftman,sun,sun2,balocchi}  Cartoix\`{a} {\em et
  al.}\cite{cartoixa} pointed out that to the first order in the momentum, the effective 
magnetic field from both the Dresselhaus and Rashba terms 
${\bf \Omega}_1=(\alpha_{\rm D}+\alpha_{\rm R})(k_y,-k_x,0)$ (with 
$z\|[111]$, $x\|[11\bar{2}]$ and $y\|[\bar{1}10]$), 
can be suppressed to zero by setting the Rashba coefficient $\alpha_{\rm R}$ to cancel with the
Dresselhaus coefficient $\alpha_{\rm D}$ and hence the DP spin relaxation becomes 
absent for all three spin components. However, 
this strict cancellation can not occur due to the existence of the cubic Dresselhaus term 
${\bf \Omega}_3=\frac{\gamma}{2\sqrt{3}}[-{\bf k}^2k_y,{\bf k}^2k_x,\sqrt{2}(3k_x^2k_y-k_y^3)]$. 
Sun {\em et al.}\cite{sun} showed that with this cubic term included, the 
cancellation occurs only for the in-plane effective 
magnetic field on a special momentum circle, which may lead to a peak in the
density or temperature dependence of the SRT. Experimentally, Balocchi {\em et al.}
\cite{balocchi} measured the gate-voltage dependence of the SRT and a two-order 
of magnitude increase of the SRT can be observed, reaching values larger than $30$\,ns 
by applying an external electric field of $50$\,kV/cm along the growth direction of the
QW. However, to the best of our knowledge, there are still 
no reports on hole spin relaxation in (111) zinc-blende QWs.

In the present work, we investigate the hole spin relaxation in $p$-type (111) GaAs QWs with only 
the lowest subband being relevant. In GaAs/AlAs QWs, the lowest hole subband is
heavy-hole (HH) like. The Dresselhaus and Rashba SOC terms of the 
lowest HH subband ${\bf \Omega}^h_{\rm D(R)}$ can be written as 
\begin{eqnarray}
{\bf
  \Omega}^h_{\rm D}&=&[\beta_x^h(3k_x^2k_y-k_y^3),\beta_y^h(k_x^3-3k_xk_y^2),\nonumber\\
&&\beta_z^h(3k_x^2k_y-k_y^3)],\label{eq1}\\
{\bf
  \Omega}^h_{\rm R}&=&[\alpha_x^h(3k_x^2k_y-k_y^3),\alpha_y^h(k_x^3-3k_xk_y^2),\nonumber\\
&&\alpha_z^h(3k_x^2k_y-k_y^3)]eE_z,
\label{eq2}
\end{eqnarray}
respectively, which are obtained by the subband L\"{o}wdin
 partitioning method.\cite{lowdin,winkler2} 
The coefficients are given in Appendix~\ref{appA}. Here, $E_z$ is the electric 
field along the growth direction. It is noted that the Rashba term ${\bf \Omega}^h_{\rm R}$ 
has not been reported in the previous literature and interestingly its 
component $\Omega^h_{{\rm R}i}$ $(i=x,y,z)$ can be strictly cancelled by the 
corresponding Dresselhaus one $\Omega^h_{{\rm D}i}$ by setting the gate voltage 
\begin{eqnarray}
E_z^i=-\beta_i^h/(e\alpha_i^h).\label{eq3} 
\end{eqnarray}
The cancellation gate voltages $E_z^x$ (about $20$\,kV/cm) and $E_z^y$ (about $22$\,kV/cm) 
are close to each other, indicating that the in-plane components of the 
effective magnetic field can be 
strongly suppressed. This cancellation leads to a strong suppression of the in-plane 
inhomogeneous broadening\cite{wu2} and hence an extremely 
long SRT for spins along 
the growth direction. It is further 
noted that although the out-of-plane component 
of the effective magnetic field of the lowest HH subband 
can also be suppressed to zero, the cancellation gate voltage $E_z^z$
 (about $10^3$\,kV/cm) 
is too large to realize. Therefore, a large anisotropy between the in-plane 
and out-of-plane spin relaxations is expected.

We also investigate the hole spin relaxation in GaAs/InP QWs 
where the lowest hole subband is light-hole
(LH) like. The Dresselhaus and Rashba SOC terms 
are then given by
\begin{eqnarray}
{\bf
  \Omega}^l_{\rm D}&=&[(\beta_x^{l1}{\bf k}^2+\beta_x^{l2}\langle
k_z^2\rangle)k_y,-(\beta_x^{l1}{\bf k}^2+\beta_x^{l2}\langle k_z^2\rangle)k_x,\nonumber\\
&&\beta_z^l(3k_x^2k_y-k_y^3)],\label{eq4}\\
{\bf
  \Omega}^l_{\rm R}&=&[\alpha_x^l{\bf
  k}^2k_y,-\alpha_x^l{\bf k}^2k_x,\alpha_z^l(3k_x^2k_y-k_y^3)]eE_z,
\label{eq5}
\end{eqnarray} 
respectively. One finds that similar to the case of electrons 
in (111) QWs,\cite{sun} the in-plane components of these 
two terms can be suppressed to zero only on a special
momentum circle.
The out-of-plane component can also be suppressed 
to zero under an unrealistic gate voltage.

We calculate the hole spin relaxation by the
kinetic spin Bloch equation (KSBE) approach,\cite{wu,wmw} 
 with all the relevant scatterings explicitly
included. As the Elliott-Yafet\cite{yafet} (EY) mechanism is
  the leading spin relaxation mechanism in hole spin
  relaxation in bulk,\cite{shen} we calculate the SRT due to both the
  DP and EY mechanisms in the present work. 
For (111) GaAs/AlAs QWs, we find that the EY mechanism is negligible
  unless the DP mechanism can be greatly suppressed. By tuning the
  gate voltage, we predict a pronounced peak in the SRT where the SRT
  can be extremely long (upto hundreds of nanoseconds). This property is favorable
for spin manipulation, storage and device design. This
peak results from the suppression of the unique inhomogeneous
  broadening\cite{wu2} in (111) QWs. The effects of hole density, temperature and impurity
density on the suppression of the spin relaxation are investigated. 
We also analyze the anisotropy of the spin relaxation. For (111)
GaAs/InP QWs, a peak is also predicted in the 
gate-voltage dependence of the SRT. However, the peak becomes 
less pronounced compared with the case in (111) GaAs/AlAs QWs. The
contribution of the EY mechanism is demonstrated to be negligible to
the predicted properties.

This paper is organized as follows. In Sec.~II, the effective
Hamiltonian of the lowest hole subband in (111) GaAs/AlAs and GaAs/InP
QWs is derived. The KSBEs are also constructed in this section.
Then in Sec.~III, we
investigate the hole spin relaxations in GaAs/AlAs and GaAs/InP QWs 
respectively. We summarize in Sec.~IV.

\section{EFFECTIVE HAMILTONIAN AND KSBEs}
We start our investigation from the $p$-type $(111)$ GaAs/AlAs and GaAs/InP
QWs. The HH and LH can be described by the $4\times 4$ Hamiltonian
\begin{equation}
  H_{4\times 4}=H_L+H_{8v8v}^{b}+H_{\epsilon}+V(z)I_4+H_EI_4,
  \label{eq6}
\end{equation}
since the contribution from the conduction and split-off bands is
marginal in the present calculation. Here, $H_L=H_L^{(0)}+H_L^{(\|)}$ is the Luttinger 
Hamiltonian\cite{luttinger} with $H_L^{(0)}$ corresponding to the part
with $k_{x,y}=0$, $H_{8v8v}^{b}$ is the Dresselhaus SOC of the HH and LH 
bands\cite{dresselhaus} and $H_{\epsilon}$ stands for the strain Hamiltonian based 
on the theory of Bir and Pikus.\cite{winkler2,bir} Their expressions are 
given in 
Appendix~\ref{appB}. $V(z)$ is the confinement applied along the growth direction within 
the infinite-depth 
well potential approximation and $I_4$ is a $4\times 4$ unit matrix. $H_E=-eE_zz$ represents 
the electric field term, which is the source of the Rashba SOC.\cite{rashba}

By solving the Schr\"odinger equation of $H_0=H_L^{(0)}+V(z)I_4+H_{\epsilon}$, the subband
energies for the HH and LH are given by
\begin{equation}
  E_{n_z}^{h,l}=\frac{{n_z}^2{\pi}^2{\hbar}^2}{2m_{z}^{h,l}a^2}\mp E_S^1-E_S^2,
  \label{eq7}
\end{equation}
where $n_z$ stands for the subband index along the growth direction, $a$
is the well width and $m_{z}^{h,l}=\frac{m_0}{\gamma_1\mp 2\gamma_3}$
represents the effective masses of the HH and LH along the $z$-axis. 
$\gamma_{1,3}$ and $m_0$ stand for the Kohn-Luttinger parameters
and the free electron mass, respectively. The contributions of the 
strain $E_S^1$ and $E_S^2$ are given in Eqs.~(\ref{eq11}) and (\ref{eq12}) in Appendix~\ref{appB}. 
It is noted that $E_S^2$ is just an energy shift
for all the HH and LH subbands and can be neglected in the present calculation.
For GaAs/AlAs QWs, the contribution of the strain is neglected due to the extremely small 
lattice mismatch and hence the lowest hole subband is HH like. However, for GaAs/InP 
QWs, due to the existence of strain, the lowest hole subband is LH like. In our
calculation, we focus on the case with only the lowest hole subband being 
relevant while other subbands have much higher energies.

\begin{figure}[bth]
\includegraphics[width=8.5cm]{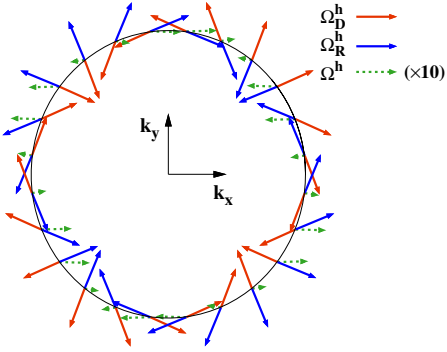}
\caption{(Color online) Schematic of the in-plane effective magnetic fields of the 
lowest HH subband in (111) GaAs/AlAs QWs around an
arbitrary equal-energy surface. Red (blue) arrows stand for 
in-plane components of the Dresselhaus (Rashba) term $\Omega_{\rm D(R)}^h$ and green 
arrows represent the sum of the Dresselhaus and Rashba terms $\Omega_h$,
with its amplitude enlarged by 10 for clarity
 at the cancellation gate voltage $E_z^y$.}
\label{fig1}
\end{figure}

We first construct the effective Hamiltonian of the lowest hole
  subband in (111) GaAs QWs by the subband L\"{o}wdin partitioning
method\cite{lowdin,winkler2} upto the third order of the in-plane
momentum. The effective Hamiltonian of the lowest hole subband in
(111) GaAs/AlAs QWs can be written as
\begin{eqnarray}
H_{\rm eff}^h&=&\varepsilon_{\bf k}^h+({\bf \Omega}^{h}_{\rm D}+{\bf \Omega}^{h}_{\rm R})\cdot
\frac{\bgreek \sigma}{2}.\label{eq8}
\end{eqnarray}
Here, $\varepsilon_{\bf k}^h=\frac{\hbar^2{\bf k}^2}{2m^h_t}$, with the 
in-plane effective mass $m_t^h$ given in Appendix~\ref{appA}, and ${\bgreek \sigma}$ 
are the Pauli matrices. The Dresselhaus and Rashba SOC terms ${\bf \Omega}^h_{\rm D(R)}$ are 
given in Eqs.~(\ref{eq1}) and (\ref{eq2}). To facilitate the understanding of the 
effective magnetic field from these SOC terms, we plot the in-plane components 
around an arbitrary equal-energy surface under the cancellation gate voltage $E_z^y$ 
[Eq.~(\ref{eq3})] in Fig.~\ref{fig1}. It is shown that 
the Dresselhaus term almost completely cancels with the Rashba one, leading to 
a strong suppression of the total in-plane effective magnetic field.

Similarly, the effective Hamiltonian of the lowest
hole subband in (111) GaAs/InP QWs reads
\begin{eqnarray}
H_{\rm eff}^l&=&\varepsilon_{\bf k}^l+({\bf \Omega}^{l}_{\rm D}+{\bf \Omega}^{l}_{\rm R})\cdot
\frac{\bgreek \sigma}{2},\label{eq9}
\end{eqnarray}
in which $\varepsilon_{\bf k}^l=\frac{\hbar^2{\bf k}^2}{2m^l_t}$ and $m_t^l$ stands for the in-plane effective mass with its form 
given in Appendix~\ref{appA}. The Dresselhaus and Rashba SOC 
terms ${\bf \Omega}^l_{\rm D(R)}$ of the LH are given in Eqs.~(\ref{eq4}) and (\ref{eq5}).

We then construct the microscopic KSBEs\cite{wu,wmw} to study the hole spin
relaxations in GaAs/AlAs and GaAs/InP QWs, respectively. The KSBEs read\cite{wu,wmw}
\begin{eqnarray}
\dot{\rho}_{{\bf k},\sigma\sigma^{\prime}}=\dot{\rho}_{{\bf k},\sigma\sigma^{\prime}}|_{\rm coh}+\dot{\rho}_{{\bf k},\sigma\sigma^{\prime}}|_{\rm scat}, 
\label{eq10}
\end{eqnarray}
where $\rho_{{\bf k},\sigma\sigma^{\prime}}$ represent the density matrix elements of
holes with the diagonal ones $\rho_{{\bf k},\sigma\sigma}\equiv f_{{\bf k}\sigma}$ $(\sigma=\pm 1/2)$
describing the distribution functions and the off-diagonal ones 
$\rho_{{\bf k},1/2\ -1/2}=\rho_{{\bf k},-1/2\ 1/2}^*\equiv\rho_{\bf k}$ denoting the spin 
coherence. $\dot{\rho}_{{\bf k},\sigma\sigma^{\prime}}|_{\rm coh}$ are the 
coherent terms describing the coherent spin 
precession due to the effective magnetic field from both the Dresselhaus and Rashba terms while the 
Hartree-Fock Coulomb interaction is neglected due to the small spin polarization.\cite{wu} Their expressions are given by 
\begin{eqnarray}
\frac{\partial f_{{\bf k},\sigma}}{\partial t}\bigg|_{\rm coh}&=&-2\sigma[\Omega_x({\bf k}){\rm Im}\rho_{\bf k}+\Omega_y({\bf k})
{\rm Re}\rho_{\bf k}],\\
\frac{\partial \rho_{{\bf k}}}{\partial t}\bigg|_{\rm coh}&=&\frac{1}{2}[i\Omega_x({\bf k})+\Omega_y({\bf k})](f_{{\bf k}1/2}
-f_{{\bf k}-1/2})\nonumber\\&&-i\Omega_z({\bf k})\rho_{\bf k}.
\end{eqnarray}
Here, $\Omega_i$ ($i=x,y,z$) are the three components of the effective magnetic field from 
both the Dresselhaus and Rashba terms. $\dot{\rho}_{{\bf k},\sigma\sigma^{\prime}}|_{\rm scat}$ are the
scattering terms including the spin conserving ones, i.e., the hole-hole ($V_{\bf q}^2$), hole-impurity
($U_{\bf q}^2$), hole-acoustic-phonon ($|M_{{\bf Q},{\rm AC}}|^2$), polar
hole-longitudinal-optical-phonon ($|M_{{\bf Q},{\rm LO}}^{\rm
  polar}|^2$) and nonpolar hole-optical-phonon 
($|M_{{\bf Q},{\lambda}}^{\rm nonp}|^2$) scatterings with
$\lambda={\rm LO/TO}$ denoting
longitudinal-optical/transverse-optical phonon and the spin-flip
  ones  due to the EY
  mechanism.\cite{yafet} Their detailed expressions can be found in
  Ref.~\onlinecite{jiang} (Note the scatterings in
  Ref.~\onlinecite{jiang} are given in bulk and one needs to transform
  them to two-dimensional cases). The scattering matrix elements $V_{\bf q}^2=(\frac{\sum_{q_z}v_Q|I(iq_z)|^2}
{\epsilon({\bf q})})^2$ and $U_{\bf
  q}^2=\sum_{q_z}[Z_iv_Q/\epsilon({\bf q})]^2|I(iq_z)|^2$, with $Z_i$
being the charge number of
the impurity (assumed to be 1 in our calculation).\cite{zhou}
$\epsilon({\bf q})=1-\sum_{q_z}v_Q|I(iq_z)|^2
\sum_{{\bf k},\sigma}\frac{f_{{\bf k}+{\bf q}\sigma}-f_{{\bf
      k}\sigma}}{\varepsilon_{{\bf k}+{\bf q}}-
\varepsilon_{\bf k}}$ stands for the screening under
random phase approximation;\cite{mahan} the bare Coulomb potential $v_Q=\frac{4\pi e^2}{Q^2}$ with $Q^2={\bf q}^2+q_z^2$;
and the form factor $|I(iq_z)|^2=\pi^4{\sin}^2y/[y^2(y^2-\pi^2)^2]$
with $y=q_za/2$. $|M_{{\bf Q},{\rm LO}}^{\rm polar}|^2=[2\pi
e^2\omega_{\rm LO}/({\bf
  q}^2+q_z^2)](\kappa_{\infty}^{-1}-\kappa_0^{-1})|I(iq_z)|^2$ with
$\omega_{\rm LO}$ denoting the LO phonon energy
and $\kappa_0$ ($\kappa_{\infty}$) being the static (optical) dielectric constant.\cite{weng} 
$|M_{{\bf Q},{\lambda}}^{\rm
  nonp}|^2=\frac{3\hbar d_0^2}{2d\omega_{\lambda}{a^2_{\rm
      GaAs}}}|I(iq_z)|^2$ for the nonpolar HH-optical-phonon
scatterings and $|M_{{\bf Q},{\lambda}}^{\rm nonp}|^2=\frac{\eta\hbar d_0^2}{2d\omega_{\lambda}{a^2_{\rm GaAs}}}|I(iq_z)|^2$  
for the nonpolar LH-optical-phonon ones with $d_0$, $d$, $a_{\rm GaAs}$, and $\omega_{\rm
  TO}$ representing the optical deformation potential,\cite{langot} the mass
density,\cite{zhou} the lattice constant of GaAs,\cite{winkler2} and
the TO phonon energy,\cite{shen} respectively. $\eta=1(5)$ for
$\lambda={\rm LO\ (TO)}$. The matrix element for the hole-acoustic phonon
scattering due to the deformation potential reads $|M_{{\bf Q},{\rm AC}}^{\rm
  def}|^2=\frac{\hbar\Xi^2Q}{2dv_{\rm sl}}|I(iq_z)|^2$ with $\Xi$
denoting the deformation potential and $v_{\rm sl}$ being the velocity of the longitudinal sound wave.\cite{zhou}
For the hole-acoustic phonon scattering due to the piezoelectric coupling, $|M_{{\bf Q},{\rm
    AC}}^{\rm pl}|^2=\frac{32\pi^2\hbar e^2e_{14}^2}{\kappa_0^2}
\frac{(3q_x^{\prime}q_y^{\prime}q_z^{\prime})^2}{dv_{\rm
    sl}Q^7}|I(iq_z)|^2$ for the longitudinal phonon and $|M_{{\bf
    Q},{\rm AC}}^{\rm pt}|^2=
\frac{32\pi^2\hbar e^2e_{14}^2}{\kappa_0^2}\frac{1}{dv_{\rm st}Q^5}[{q_x^{\prime}}^2{q_y^{\prime}}^2+{q_y^{\prime}}^2{q_z^{\prime}}^2+
{q_z^{\prime}}^2{q_x^{\prime}}^2-\frac{(3q_x^{\prime}q_y^{\prime}q_z^{\prime})^2}{Q^2}]|I(iq_z)|^2$
for the transverse phonon, where
$q_x^{\prime}=\frac{1}{\sqrt{6}}q_x-\frac{1}{\sqrt{2}}q_y+\frac{1}{\sqrt{3}}q_z$, $q_y^{\prime}=\frac{1}{\sqrt{6}}q_x+\frac{1}{\sqrt{2}}q_y+
\frac{1}{\sqrt{3}}q_z$ and $q_z^{\prime}=-\frac{2}{\sqrt{6}}q_x+\frac{1}{\sqrt{3}}q_z$. $v_{\rm st}$
represents the velocity of the transverse sound wave and $e_{14}$ is
the piezoelectric constant.\cite{zhou} The spin mixing $\hat{\Lambda}_{{\bf k},{\bf k^{\prime}}}$ for the
lowest HH and LH subbands in the spin-flip scattering terms are detailed in Appendix~\ref{appC}.

\begin{table}
\caption{Parameters used in the calculation (from
  Ref.~\onlinecite{winkler2} unless otherwise specified)}
\begin{tabular}{llllll}
  \hline
  \hline 
  $\gamma_1$ &6.85  &$\gamma_2$   &2.10  &$\gamma_3$   &2.90\\
  $b_{41}$    &$-$81.93 eV\r{A}$^3$  &$b_{42}$    &1.47 eV\r{A}$^3$ &$b_{51}$    &0.49 eV\r{A}$^3$\\
  $b_{52}$    &$-$0.98 eV\r{A}$^3$   &$D_u^{\prime}$ &4.67 eV   &$a_{\rm GaAs}$ &5.65 \r{A}\\
  $a_{\rm AlAs}$  &5.66 \r{A} &$a_{\rm InP}$ &5.87 \r{A} &$C_{11}$  &11.81\\
  $C_{12}$ &5.32  &$C_{44}$ &5.94 &$d$  &5.31 g/cm$^3$\\
  $d_0$ &48 eV$^{\rm a}$&$\omega_{\rm LO}$ &35.4 meV$^{\rm b}$ &$\omega_{\rm TO}$
  &33.2 meV$^{\rm b}$\\
  $e_{14}$ &1.41$\times 10^9$ V/m$^{\rm c}$  &$\kappa_{\infty}$   &10.8$^{\rm c}$ &$\Xi$ &8.5 eV$^{\rm c}$\\
  $v_{\rm sl}$ &5.29$\times 10^3$ m/s$^{\rm c}$ &$v_{\rm st}$ &2.48$\times 10^3$
  m/s$^{\rm c}$ &$\kappa_0$ &12.9$^{\rm c}$\\
\hline
\hline
\end{tabular}
\begin{tabular}{llll}
\hspace{-4.4cm}$^{\rm a}$Reference \onlinecite{langot} & & &\\
\hspace{-4.4cm}$^{\rm b}$Reference \onlinecite{weng} & & &\\
\hspace{-4.4cm}$^{\rm c}$Reference \onlinecite{zhou} & & &\\
\end{tabular}
\end{table}

\section{NUMERICAL RESULTS}
We present our results obtained by numerically solving
the KSBEs. All the parameters used in our calculation are listed in
Table~I. The initial spin polarization is along the QW growth direction except otherwise
specified. We set the initial spin polarization as
$2.5$~\%. Note that the SRT is insensitive to the small spin polarization.\cite{weng} It is
noted that the system is always in the strong scattering limit in
the present investigation.

\begin{figure}[bth]
\includegraphics[width=7.5cm]{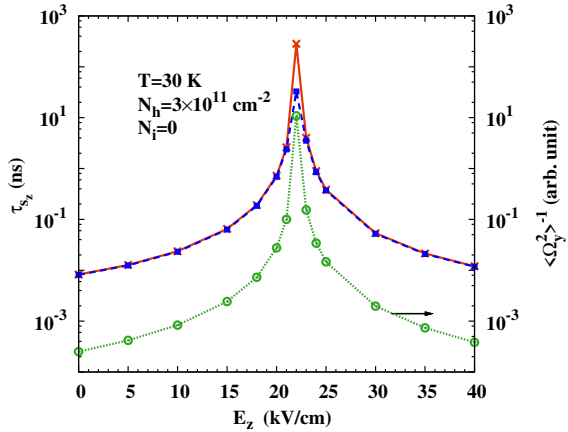}
\caption{(Color online) HH SRT along the $z$-axis $\tau_{s_z}$ and the inverse of the inhomogeneous
  broadening ${\langle\Omega^2_{y}\rangle}^{-1}$ of the lowest hole
  subband in (111) GaAs/AlAs QWs against the gate voltage
  $E_z$. Red solid (blue dashed) curve represents
 the SRT due to the DP (both DP and EY)
  mechanism. The temperature $T=30$\,K, the hole density $N_h=3\times
  10^{11}$\,cm$^{-2}$ and the impurity density $N_i=0$. Note the scale
  of ${\langle\Omega^2_{y}\rangle}^{-1}$ is on the right-hand side of
  the frame.}
\label{fig2}
\end{figure}

\subsection{HH spin relaxation in GaAs/AlAs QWs}
Before a full investigation by numerically solving the KSBEs, we start
from an analytical analysis of the DP spin relaxation of the lowest
hole subband, which is  HH like, in (111) GaAs/AlAs QWs. In the strong
scattering limit, the evolution of spin polarization
along the $z$-axis can be approximated by\cite{pikus}
\begin{eqnarray}
\dot{S_z}=-\tau_p^{*}[S_z\langle\Omega_x^2+\Omega_y^2\rangle-S_x\langle\Omega_x\Omega_z\rangle-S_y\langle
\Omega_y\Omega_z\rangle],
\end{eqnarray}
with $\tau_p^{*}$ standing for the effective momentum scattering time
contributed by the hole-hole, hole-impurity and hole-phonon
scatterings.\cite{weng,wu2} The evolutions
of the in-plane components $S_x$ and $S_y$ can be obtained by index
permutation. It is noted that unlike (001) III-V zinc-blende QWs where
 $\Omega_z$ is absent and the inhomogeneous
broadening of the spin polarization along the $z$-axis is determined by
$\langle\Omega_{\perp}^2\rangle=\langle\Omega_x^2+\Omega_y^2\rangle$,\cite{wu}
for ($111$) QWs, $\Omega_z$ and $\langle\Omega_x\Omega_z\rangle$
are nonzero under the effective magnetic field from both the Dresselhaus and
Rashba terms given in Eqs.~(\ref{eq1}) and (\ref{eq2}). This makes
the evolutions more complex compared with (001) QWs, i.e., 
\begin{eqnarray}
\dot{S_z}&=&-\tau_p^{*}[S_z\langle\Omega_x^2+\Omega_y^2\rangle-S_x\langle\Omega_x\Omega_z\rangle],\label{eq19}\\
\dot{S_x}&=&-\tau_p^{*}[S_x\langle\Omega_y^2+\Omega_z^2\rangle-S_z\langle\Omega_x\Omega_z\rangle],\label{eq20}\\
\dot{S_y}&=&-\tau_p^{*}[S_y\langle\Omega_x^2+\Omega_z^2\rangle].\label{eq21}
\end{eqnarray}
The exact solutions are given in Appendix~\ref{appD}. Specially by setting the
initial spin polarization along the $z$-axis as $S_z(0)$, we obtain
$S_z=S_z(0)e^{-t/\tau_{s_z}}$ approximately by considering that
$\langle\Omega_{x}^2\rangle$, $\langle\Omega_{y}^2\rangle$ and
$\langle\Omega_x\Omega_z\rangle$ are much smaller than
$\langle\Omega_z^2\rangle$ in our investigation.
The SRT along the $z$-axis reads
\begin{eqnarray}
\tau_{s_z}&=&{\tau_p^{*}}^{-1}{\langle\Omega_y^2\rangle}^{-1}\label{eq13},
\end{eqnarray}
with ${\langle\Omega_y^2\rangle}$ being the inhomogeneous broadening.
This inhomogeneous broadening is unique in (111) QWs and has not been 
reported in the literature.

We then numerically calculate the spin relaxation of the lowest hole
subband, which is HH like, in (111)
GaAs/AlAs QWs where the well width is chosen to be $8$\,nm. We
  first investigate the SRT due to the DP mechanism and  plot the
  gate-voltage dependence of the SRT along the $z$-axis
 at $T=30$\,K with the hole density $N_h=3\times
10^{11}$\,cm$^{-2}$ by red solid curve in Fig.~\ref{fig2}.
It is seen that the SRT can be effectively modulated by the gate
voltage and a peak appears at $E_z\approx 22$\,kV/cm. Under this
gate voltage, the SRT is as long as $283$\,ns which is about four orders of
magnitude larger than the one without any bias. This peak results from
the suppression of the inhomogeneous broadening
$\langle\Omega_y^2\rangle$ as shown by green 
dotted curve in Fig.~\ref{fig2}. This suppression originates from the 
cancellation of the $y$-component of the effective magnetic field $\Omega_y$ [Eq.~(\ref{eq3})].

\begin{figure}[bth]
\centering
\includegraphics[width=4.2cm]{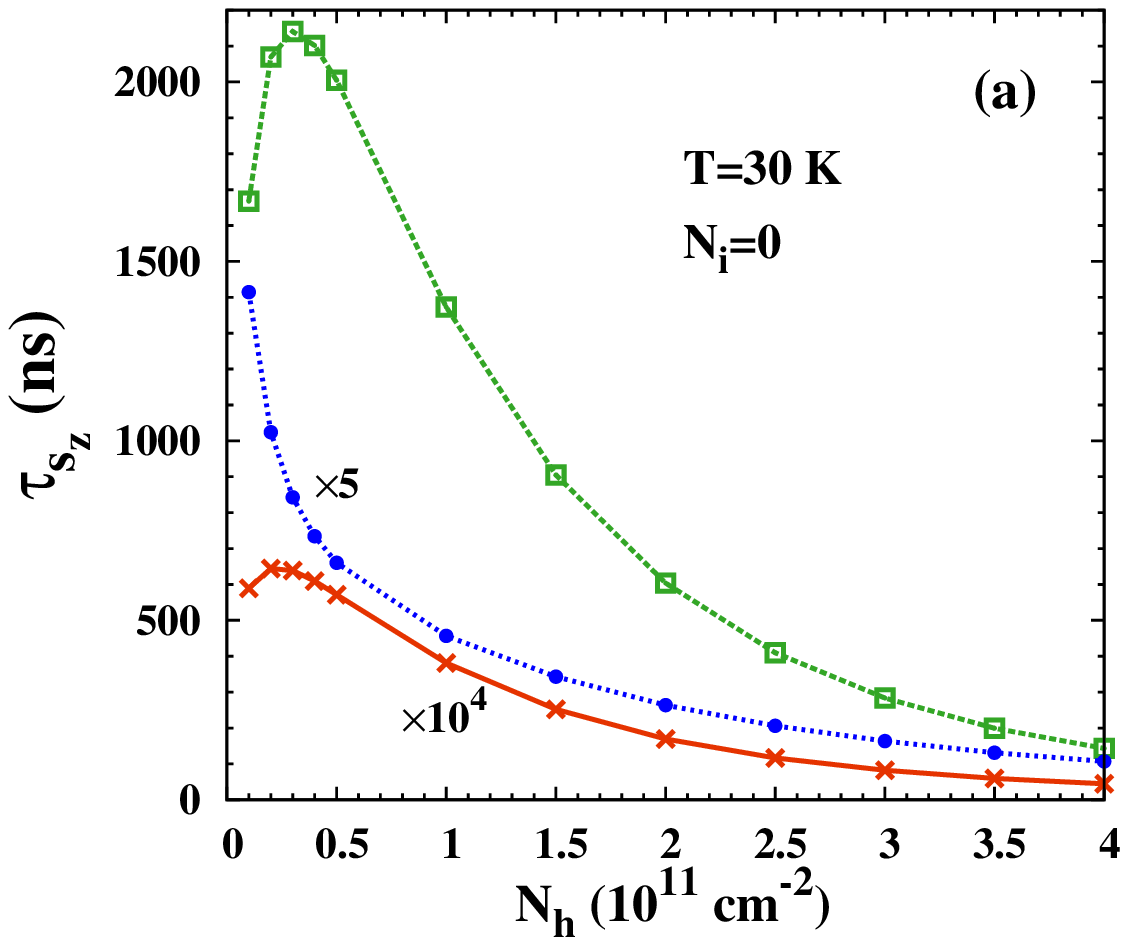}
\includegraphics[width=4.2cm]{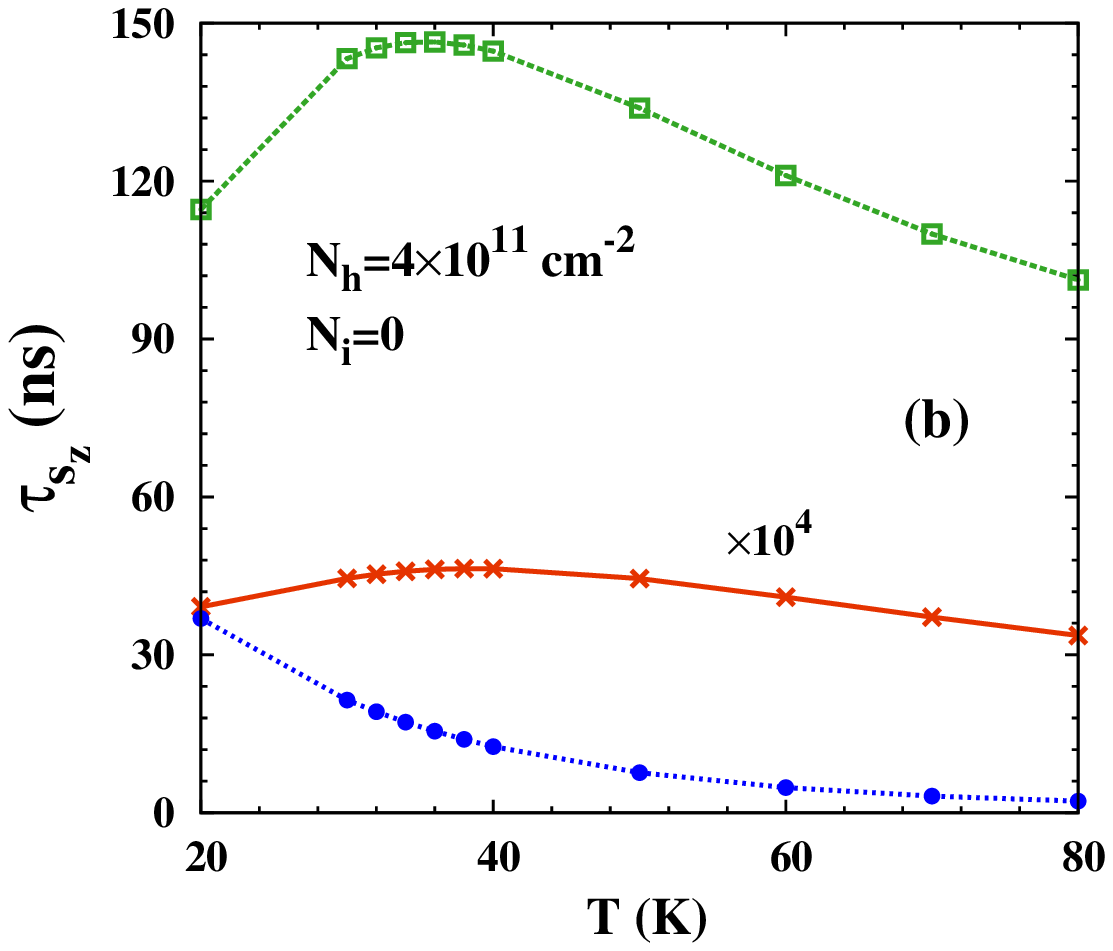}
\includegraphics[width=4.2cm]{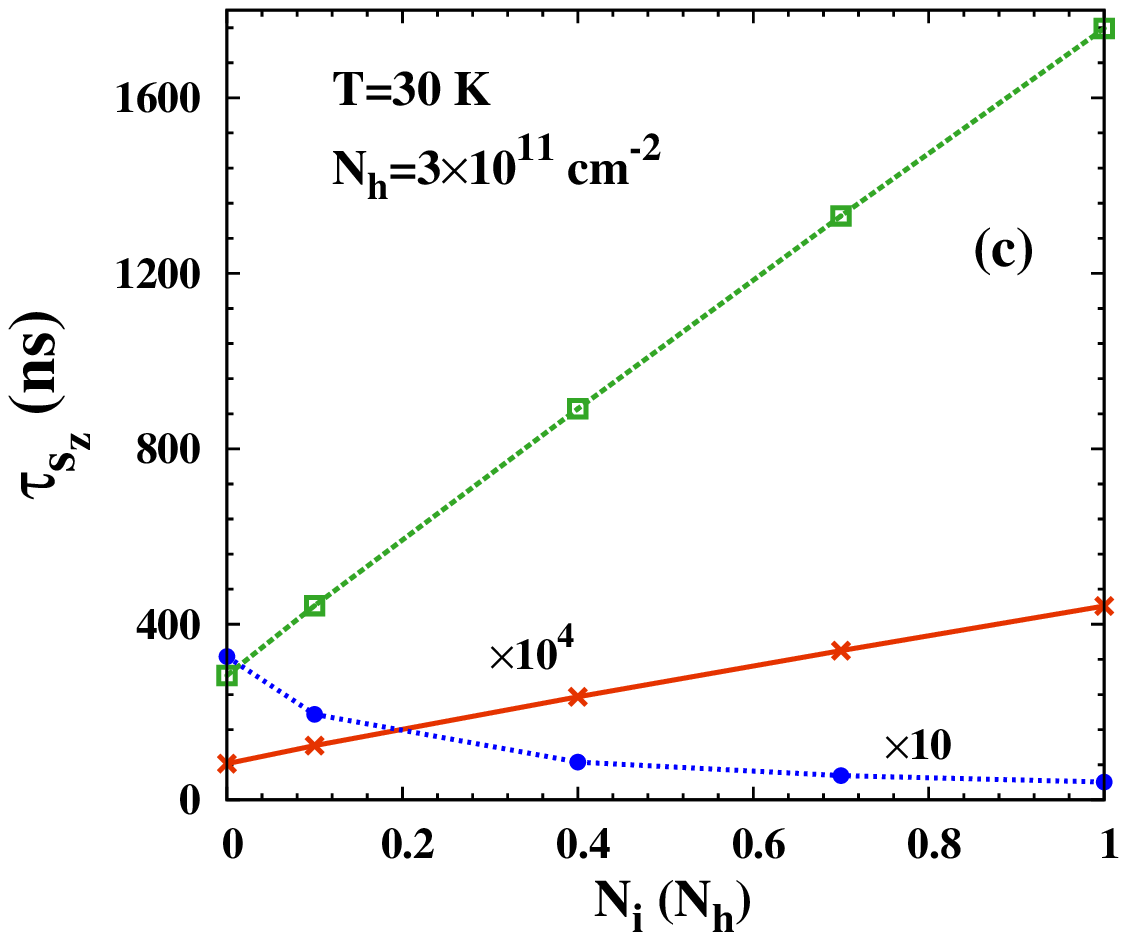}
\includegraphics[width=4.2cm]{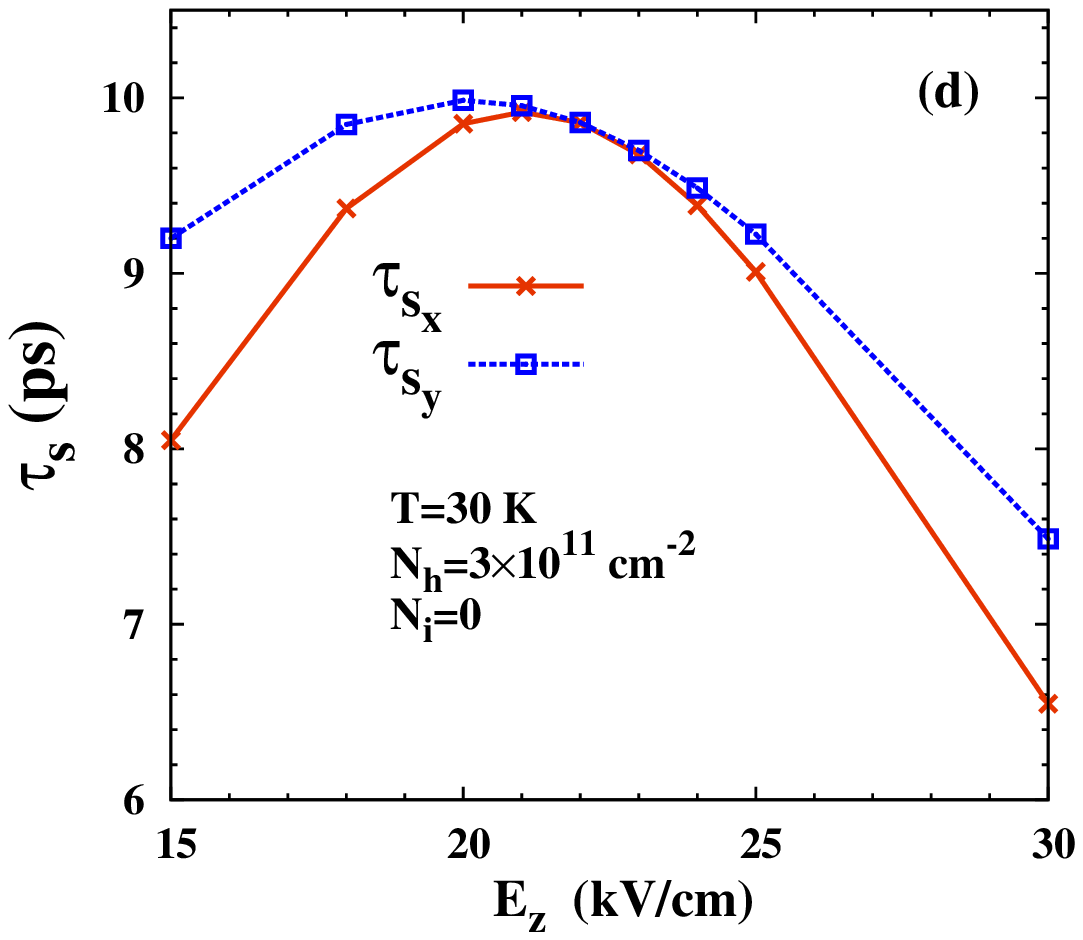}
\caption{(Color online) HH SRT along the $z$-axis $\tau_{s_z}$
 of the lowest hole subband in
  (111) GaAs/AlAs QWs as a function of (a) the hole density with
  temperature $T=30$\,K and impurity density $N_i=0$; 
(b) temperature with the hole density $N_h=4\times
  10^{11}$\,cm$^{-2}$ and $N_i=0$; and (c)  impurity
  density with $T=30$\,K and $N_h=3\times
  10^{11}$\,cm$^{-2}$. Red curves with $\times$ stand for the
  SRTs due to the DP mechanism under zero bias. Green curves with $\Box$
 (blue ones with $\bullet$) represent the SRTs due to the DP (both DP and
  EY) mechanism at the cancellation gate voltage $E_z=22$\,kV/cm. (d)
  HH SRT along the $x$ ($y$)-axis $\tau_{s_x}$ ($\tau_{s_y}$) against the
  gate voltage $E_z$ with $T=30$\,K, $N_h=3\times 10^{11}$\,cm$^{-2}$
  and $N_i=0$.}
\label{fig3}
\end{figure}

Since the suppression of the DP spin relaxation is determined by the
cancellation of the $y$-component of the effective magnetic 
field $\Omega_y$ [Eq.~(\ref{eq3})], it is robust against the hole density, temperature
and impurity density. In Fig.~\ref{fig3}(a)-(c),
 we plot the hole density, temperature
and impurity density dependences of the SRT under zero bias and the
cancellation gate voltage $E_z=22$\,kV/cm as red curves with $\times$
and green ones with $\Box$, respectively. It is shown that the SRT at
the cancellation gate voltage is always four orders of
magnitude larger than the one under zero bias. In addition, we also
observe peaks in the hole density and temperature dependences with the peak
locations fixed under different gate voltages. The peak in the density
dependence of the SRT has been theoretically predicted by Jiang and Wu\cite{jiang} and experimentally 
confirmed\cite{kraub,bub,han,han2} recently, which is
attributed to the crossover from the nondegenerate to degenerate
limit. The peak in the temperature dependence of the SRT was predicted by
Zhou {\em et al.}\cite{zhou} and experimentally realized by Leyland {\em et al.},\cite{leyland}
Ruan {\em et al.},\cite{ruan} and Han {\em et al.},\cite{han} which is
solely caused by the Coulomb scattering.\cite{zhou} The location
of the peak is insensitive to the gate voltage just the same as the case for
electrons in $(111)$ QWs.\cite{sun}

It is further noted that the EY mechanism is usually the major
mechanism in hole spin relaxation in bulk GaAs.\cite{shen} In order to rule out
the influence of the EY mechanism to the predicted phenomena, we 
plot in Fig.~\ref{fig2} the gate-voltage
dependence of the SRT due to both the
DP and EY mechanisms as the blue dashed curve. It
is seen that the EY mechanism can be important only around the
cancellation gate voltage where the DP spin relaxation is greatly
suppressed. Away from the cancellation gate voltage, the contribution of
the EY mechanism is negligible since the spin mixing between the HH
and LH is extremely small due to the large energy splitting between
the HH and LH subbands in QWs with small well width. 
Moreover, since the EY mechanism is important
at the cancellation gate voltage, we also plot in
Fig.~\ref{fig3}(a)(c) the hole density,
temperature and impurity density dependences of the SRT due to both
the EY and DP mechanisms under this gate voltage (blue curves with
$\bullet$). We find that the peaks in the hole density and
temperature dependences disappear since the SRT caused by the EY
mechanism decreases with the hole density and temperature.\cite{semi,wu,shen} 
Moreover, the SRT due to the EY mechanism also decreases with the impurity density.\cite{semi,wu,shen}
However, away from the cancellation field, the spin relaxation is
determined by the DP mechanism only and the peaks still exist in the
corresponding temperature and density dependences.

Finally, we address the anisotropy of the spin relaxation with respect
to the spin polarization direction. The SRT along the $z$-axis
$\tau_{s_z}$ is given in Eq.~(\ref{eq13}). Similarly, one obtains the
SRTs along the $x$- and $y$-axes  as
\begin{eqnarray}
\tau_{s_x}&=&{\tau_p^{*}}^{-1}{\langle\Omega_x^2+\Omega_y^2+\Omega_z^2\rangle}^{-1},\label{eq14}\\
\tau_{s_y}&=&{\tau_p^{*}}^{-1}{\langle\Omega_x^2+\Omega_z^2\rangle}^{-1},\label{eq15}
\end{eqnarray}
by setting the initial spin polarizations along the $x$- and $y$-axes
correspondingly. By comparing Eq.~(\ref{eq13}) with Eqs.~(\ref{eq14}) and
(\ref{eq15}), one finds a strong anisotropy between the out-of- and in-plane spin
relaxation, since $\Omega_z$ is about one order of magnitude larger than the in-plane
components $\Omega_{x}$ and $\Omega_{y}$ in (111) QWs with small well width.
In addition, a marginal anisotropy also arises between
the in-plane spin relaxations from Eqs.~(\ref{eq14}) and
(\ref{eq15}). The numerical result of the SRT along the $x$ 
($y$)-axis $\tau_{s_x}$
  ($\tau_{s_y}$) is shown in Fig.~\ref{fig3}(d) by red curve with
  $\times$ (blue one with $\Box$). One finds from the figure that
both $\tau_{s_x}$ and $\tau_{s_y}$ are
  about two orders of magnitude smaller than $\tau_{s_z}$.

\begin{figure}[bth]
\includegraphics[width=6.cm]{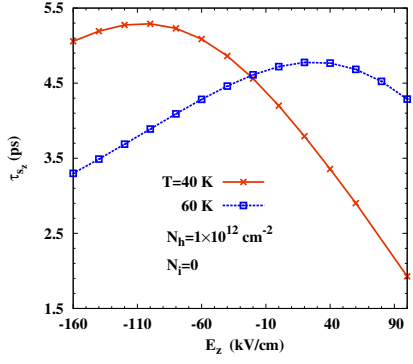}
\caption{(Color online) LH SRT along the $z$-axis $\tau_{s_z}$ 
of the lowest hole subband in
  (111) GaAs/InP QWs against the gate voltage $E_z$. Red curve 
with $\times$ (blue
  one with $\Box$) stands for the case with 
 temperature $T=40$ ($60$)\ K. The hole
  density $N_h=1\times 10^{12}$\,cm$^{-2}$ and impurity density $N_i=0$.}
\label{fig4}
\end{figure}

\subsection{LH spin relaxation in GaAs/InP QWs}
We also investigate the spin relaxation of the lowest hole subband,
which is LH like, in (111)
GaAs/InP QWs with the well width $a=20$\,nm. The gate-voltage
dependence of the SRT along the $z$-axis due to the DP mechanism under different 
temperatures is shown in Fig.~\ref{fig4} and a peak is also
observed. We find that the peak becomes less pronounced than
that in GaAs/AlAs QWs since the cancellation only occurs on a special
momentum circle in GaAs/InP QWs.\cite{sun}
In addition, when the temperature increases, 
the peak location changes, similar to the case for
electrons in (111) GaAs QWs.\cite{sun}  It is noted that the 
contribution of the EY
mechanism is in the order of 10~ns and is therefore always negligible 
in this case.

\section{SUMMARY}
In summary, we have investigated the hole spin relaxations in $p$-type
(111) GaAs/AlAs and GaAs/InP QWs with only the lowest hole subband
being relevant. The lowest hole subband is HH like in the former case
while LH like in the latter one. In both systems, we utilize the subband
L\"{o}wdin perturbation method to obtain the effective Hamiltonian
including the Dresselhaus and Rashba SOCs. Under a proper gate
voltage, the total in-plane effective magnetic
field in (111) GaAs/AlAs QWs can be strongly suppressed in the whole
momentum space, while the one in (111) GaAs/InP QWs can only be
 suppressed to zero on a
special momentum circle. Therefore, tuning the gate voltage is an
efficient way to modulate the hole spin relaxation in (111) QWs.

We calculate the hole spin relaxation due to the DP and EY
 mechanisms by numerically solving the
KSBEs with all the relevant scatterings explicitly  included.
For (111) GaAs/AlAs QWs, the contribution of the EY mechanism
is negligible unless the DP mechanism is strongly suppressed.
In addition, we predict a pronounced peak  in the gate
voltage-dependence of the HH SRT where an extremely long SRT (upto
hundreds of nanoseconds) can be reached. This peak comes from the
suppression of the unique inhomogeneous broadening in (111) GaAs/AlAs QWs,
specifically, the cancellation of the $y$-component of the effective magnetic
field. The suppression of the spin relaxation is
robust against the hole density, temperature and
impurity density.  A strong anisotropy between the out-of- and
in-plane spin relaxations is also addressed. In (111) GaAs/InP QWs, a peak
is also predicted in the gate-voltage dependence of the LH SRT.
However, the peak becomes less pronounced compared 
with the case in (111) GaAs/AlAs QWs since the
cancellation only occurs on a special momentum circle in GaAs/InP QWs. The contribution
of the EY mechanism is always negligible in this case. This
investigation suggests that the HH in (111) GaAs/AlAs QWs offers unique
properties for the spintronic devices and calls for experimental investigations.

\begin{acknowledgments}
This work was supported by the National Basic Research Program of
China under Grant No.\ 2012CB922002 and
the National Natural Science Foundation of China under Grant No.\
10725417.
\end{acknowledgments}

\begin{appendix}
\section{EFFECTIVE MASS AND SPIN-ORBIT COUPLING COEFFICIENTS}\label{appA}

The in-plane effective mass $m_t^h$ in Eq.~(\ref{eq8}) and the Dresselhaus (Rashba) strengths
$\beta_{i}^h$ ($\alpha_i^h$) $(i=x,y,z)$ of the lowest hole subband,
which is HH like, in Eqs.~(\ref{eq1}) and (\ref{eq2}) are given by
\begin{equation}
\frac{1}{m_t^h}=\frac{\gamma_1+\gamma_3}{m_0}-\sum_{n_z\ne
  1}\frac{2B_1^2Q_{1n_z}^2}{\hbar^2\Delta_{n_z1}^{lh}}-\frac{2C_1^2}{\hbar^2\Delta_{11}^{lh}},
\end{equation}
\begin{eqnarray}
\beta_x^h&=&\frac{b_{42}}{\sqrt{3}}-\frac{4(C_1B_3+B_2C_3)}{\Delta_{11}^{lh}}\nonumber\\
&&\mbox{}+\sum_{n_z\ne 1}\frac{4(B_1C_4+2C_2B_2)Q_{1n_z}^2}{\Delta_{n_z1}^{lh}},
\end{eqnarray}
\begin{eqnarray}
\beta_y^h&=&\frac{-b_{51}+b_{52}}{\sqrt{3}}+\frac{4(C_1B_3-B_2C_3)}{\Delta_{11}^{lh}}\nonumber\\
&&\mbox{}+\sum_{n_z\ne 1}\frac{4(2C_2B_2-B_1C_4)Q_{1n_z}^2}{\Delta_{n_z1}^{lh}},
\end{eqnarray}
\begin{eqnarray}
\beta_z^h&=&-\frac{\sqrt{6}}{2}b_{41}-\frac{23\sqrt{6}}{24}b_{42}+\frac{4(C_3B_3-C_1B_2)}{\Delta_{11}^{lh}}\nonumber\\
&&\mbox{}+\sum_{n_z\ne 1}\frac{4(B_1C_2-2B_2C_4)Q_{1n_z}^2}{\Delta_{n_z1}^{lh}},
\end{eqnarray}
\begin{eqnarray}
\alpha_x^h&=&\sum_{n_z\ne
  1}\frac{4P_{1n_z}Q_{1n_z}}{\Delta_{n_z1}^{lh}\Delta_{n_z1}^{hh}}(B_1B_3-C_1C_4n_z^2+2B_2^2\nonumber\\
&&\mbox{}-C_2C_3n_z^2)+\sum_{n_z\ne 1}\frac{4P_{1n_z}Q_{1n_z}}{\Delta_{11}^{lh}}(\frac{1}{\Delta_{n_z1}^{lh}}-\frac{1}{\Delta_{n_z1}^{hh}})\nonumber\\
&&\mbox{}\times(B_1B_3-C_1C_4+2B_2^2-C_2C_3),
\end{eqnarray}
\begin{eqnarray}
\alpha_y^h&=&\sum_{n_z\ne
  1}\frac{4P_{1n_z}Q_{1n_z}}{\Delta_{n_z1}^{lh}\Delta_{n_z1}^{hh}}(B_1B_3-C_1C_4n_z^2-2B_2^2\nonumber\\
&&\mbox{}+C_2C_3n_z^2)+\sum_{n_z\ne 1}\frac{4P_{1n_z}Q_{1n_z}}{\Delta_{11}^{lh}}(\frac{1}{\Delta_{n_z1}^{lh}}-\frac{1}{\Delta_{n_z1}^{hh}})\nonumber\\
&&\mbox{}\times(B_1B_3-C_1C_4-2B_2^2+C_2C_3),
\end{eqnarray}
\begin{eqnarray}
\alpha_z^h&=&\sum_{n_z\ne
  1}\frac{4P_{1n_z}Q_{1n_z}}{\Delta_{n_z1}^{lh}\Delta_{n_z1}^{hh}}(B_1B_2-C_1C_2n_z^2-2B_2B_3\nonumber\\
&&\mbox{}+C_3C_4n_z^2)+\sum_{n_z\ne 1}\frac{4P_{1n_z}Q_{1n_z}}{\Delta_{11}^{lh}}(\frac{1}{\Delta_{n_z1}^{lh}}-\frac{1}{\Delta_{n_z1}^{hh}})\nonumber\\
&&\mbox{}\times(B_1B_2-C_1C_2-2B_2B_3+C_3C_4).
\end{eqnarray}

The in-plane effective mass $m_t^l$ in Eq.~(\ref{eq9}), the Dresselhaus strengths
$\beta_x^{l1,l2}$, $\beta_z^l$ and the Rashba strengths
$\alpha_{x,z}^l$ of the lowest hole subband, which is LH like,
in Eqs.~(\ref{eq4}) and (\ref{eq5}) read
\begin{eqnarray}
\frac{1}{m_t^l}&=&\frac{\gamma_1-\gamma_3}{m_0}-\frac{2(C_1^2+C_3^2)}{\hbar^2\Delta_{11}^{hl}}
-\sum_{n_z\ne 1}\frac{2}{\hbar^2\Delta_{n_z1}^{hl}}\nonumber\\&&\mbox{}\times(B_1^2+4B_2^2)Q_{1n_z}^2,
\end{eqnarray}
\begin{eqnarray}
\beta_x^{l1}&=&\frac{4b_{41}+7b_{42}}{4\sqrt{3}}-\frac{\sqrt{3}}{2}(b_{51}+b_{52})
-\frac{4}{\Delta_{11}^{hl}}(B_2C_3\nonumber\\
&&\mbox{}-C_1B_3)+\sum_{n_z\ne 1}\frac{4(-B_1C_4+2C_2B_2)Q_{1n_z}^2}{\Delta_{n_z1}^{hl}},\\
\beta_x^{l2}&=&-\frac{4\sqrt{3}b_{41}}{3}-\frac{7\sqrt{3}b_{42}}{3}-\sqrt{3}b_{52},\\
\beta_z^l&=&-\frac{\sqrt{6}b_{41}}{6}-\frac{13\sqrt{6}b_{42}}{24}-\frac{4(C_1B_2+C_3B_3)}{\Delta_{11}^{hl}}\nonumber\\
&&\mbox{}+\sum_{n_z\ne 1}\frac{4(B_1C_2+2B_2C_4)Q_{1n_z}^2}{\Delta_{n_z1}^{hl}},
\end{eqnarray}
\begin{eqnarray}
\alpha_x^l&=&\sum_{n_z\ne
  1}\frac{4P_{1n_z}Q_{1n_z}}{\Delta_{n_z1}^{ll}\Delta_{n_z1}^{hl}}(B_1B_3-C_1C_4n_z^2-2B_2^2\nonumber\\
&&\mbox{}+C_2C_3n_z^2)+\sum_{n_z\ne
  1}\frac{4P_{1n_z}Q_{1n_z}}{\Delta_{11}^{hl}}(\frac{1}{\Delta_{n_z1}^{hl}}-\frac{1}{\Delta_{n_z1}^{ll}})\nonumber\\
&&\mbox{}\times(B_1B_3-C_1C_4-2B_2^2+C_2C_3),\\
\alpha_z^l&=&-\sum_{n_z}\frac{4W_1P_{1n_z}Q_{1n_z}}{\Delta_{11}^{hl}\Delta_{n_z1}^{hl}}
-\sum_{n_z\ne
  1}\frac{4W_1P_{1n_z}Q_{1n_z}n_z^2}{\Delta_{n_z1}^{{ll}^2}}\nonumber\\
&&\mbox{}-\sum_{n_z\ne 1}\frac{4P_{1n_z}Q_{1n_z}}{\Delta_{n_z1}^{ll}\Delta_{11}^{hl}}(-B_1B_2+C_1C_2-2B_2B_3\nonumber\\
&&\mbox{}+C_3C_4)+\sum_{n_z\ne 1}\frac{4P_{1n_z}Q_{n_z1}}{\Delta_{n_z1}^{ll}\Delta_{n_z1}^{hl}}(B_1B_2-C_1C_2n_z^2\nonumber\\
&&\mbox{}+2B_2B_3-C_3C_4n_z^2).
\end{eqnarray}

In above equations,  $\gamma_i$ are Kohn-Luttinger parameters; $b_{41,42,51,52}$
stand for the coefficients of the Dresselhaus SOC of the valence band;
$\Delta_{n_1n_2}^{hh}$ ($\Delta_{n_1n_2}^{ll}$) is the energy difference
between $n_1$-th and $n_2$-th subbands of HH (LH); $\Delta_{n_1n_2}^{hl}$ is the
energy difference between $n_1$-th HH and $n_2$-th LH subbands.
$B_1=\frac{\sqrt{3}\hbar^2(2\gamma_2+\gamma_3)}{3m_0}$,
$B_2=\frac{\sqrt{6}\hbar^2(-\gamma_2+\gamma_3)}{6m_0}$, and
$B_3=\frac{\sqrt{3}\hbar^2(\gamma_2+2\gamma_3)}{6m_0}$.
$C_1=(b_{41}+\frac{9}{4}b_{42}-\frac{1}{2}b_{52})\frac{\pi^2}{a^2}$,
$C_2=\frac{\sqrt{2}}{4}(b_{41}+\frac{9}{4}b_{42}-b_{51}+b_{52})$,
$C_3=\frac{\sqrt{2}\pi^2}{4a^2}(2b_{42}+b_{52})$,
and
$C_4=\frac{1}{4}(b_{42}+b_{51}-b_{52})$.
$P_{n_1n_2}=\frac{4an_1n_2[(-1)^{n_1+n_2}-1]}{(n_1^2-n_2^2)^2\pi^2}(1-\delta_{n_1,n_2})$,
$Q_{n_1n_2}=\frac{2n_1n_2[(-1)^{n_1+n_2}-1]}{(n_1^2-n_2^2)a}(\delta_{n_1,n_2}-1)$,
and $W_1=\frac{3\sqrt{2}\pi^2}{a^2}(\frac{1}{6}b_{41}+\frac{7}{24}b_{42}+\frac{1}{4}b_{51}-\frac{1}{4}b_{52})
(\frac{2}{3}b_{41}+\frac{7}{6}b_{42}+\frac{1}{2}b_{52})$.

\section{LUTTINGER HAMILTONIAN FOR (111)-ORIENTED III-V ZINC-BLENDE CRYSTAL}\label{appB}
The Luttinger Hamiltonian $H_L$, the Dresselhaus SOC $H_{8v8v}^{b}$
and the strain Hamiltonian $H_{\epsilon}$ in Eq.~(\ref{eq6}) are given
in the following. The Luttinger Hamiltonian $H_L$ can be written as\cite{luttinger}
\begin{eqnarray}
H_L=\left(\begin{array}{cccc}
F & H & I & 0 \\
H^* & G & 0 & I \\
I^* & 0 & G & -H \\
0 & I^* & -H^* & F 
\end{array}\right)
\end{eqnarray}
in the basis of the eigenstates of $J_z$ with eigenvalues $\frac{3}{2}, \frac{1}{2}, -\frac{1}{2}$, and $-\frac{3}{2}$ 
in sequence, where $F=\frac{\hbar^2}{2m_0}[(\gamma_1+\gamma_3)(k_x^2+k_y^2)+(\gamma_1-2\gamma_3)k_z^2]$, 
$G=\frac{\hbar^2}{2m_0}[(\gamma_1-\gamma_3)(k_x^2+k_y^2)+(\gamma_1+2\gamma_3)k_z^2]$, 
$H=-\frac{\hbar^2}{2m_0}[\frac{\sqrt{6}}{3}(-\gamma_2+\gamma_3)k_+^2+\frac{2\sqrt{3}}{3}(2\gamma_2+\gamma_3)k_-k_z]$,
and
$I=-\frac{\hbar^2}{2m_0}[\frac{\sqrt{3}}{3}(\gamma_2+2\gamma_3)k_-^2+\frac{2\sqrt{6}}{3}(-\gamma_2+\gamma_3)k_+k_z]$,
with $k_{\pm}=k_x\pm ik_y$. $H_L^{(0)}=\frac{\hbar^2k_z^2}{2m_0}{\rm diag}(\gamma_1-2\gamma_3,\gamma_1+2\gamma_3,\gamma_1+2\gamma_3,
\gamma_1-2\gamma_3)$ corresponds to the part of the Luttinger Hamiltonian $H_L$ by setting $k_{x,y}=0$. 

The Dresselhaus SOC $H_{8v8v}^{b}$ is given by\cite{winkler2}
\begin{eqnarray}
H_{8v8v}^b&=&-b_{41}(\{k_x^{\prime},{k_y^{\prime}}^2-{k_z^{\prime}}^2\}J_x^{\prime}+{\rm cp})-b_{42}(\{k_x^{\prime},{k_y^{\prime}}^2\nonumber\\
&&\mbox{}-{k_z^{\prime}}^2\}{J_x^{\prime}}^3+{\rm cp})-b_{51}(\{k_x^{\prime},{k_y^{\prime}}^2+{k_z^{\prime}}^2\}\{J_x^{\prime},{J_y^{\prime}}^2\nonumber\\
&&\mbox{}-{J_z^{\prime}}^2\}+{\rm
  cp})-b_{52}({k_x^{\prime}}^3\{J_x^{\prime},{J_y^{\prime}}^2-{J_z^{\prime}}^2\}+{\rm cp}),\nonumber\\
\end{eqnarray}
where
$k_x^{\prime}=\frac{1}{\sqrt{6}}k_x-\frac{1}{\sqrt{2}}k_y+\frac{1}{\sqrt{3}}k_z$,
$k_y^{\prime}=\frac{1}{\sqrt{6}}k_x+\frac{1}{\sqrt{2}}k_y+\frac{1}{\sqrt{3}}k_z$,
$k_z^{\prime}=-\frac{2}{\sqrt{6}}k_x+\frac{1}{\sqrt{3}}k_z$,
$J_x^{\prime}=\frac{1}{\sqrt{6}}J_x-\frac{1}{\sqrt{2}}J_y+\frac{1}{\sqrt{3}}J_z$,
$J_y^{\prime}=\frac{1}{\sqrt{6}}J_x+\frac{1}{\sqrt{2}}J_y+\frac{1}{\sqrt{3}}J_z$,
and $J_z^{\prime}=-\frac{2}{\sqrt{6}}J_x+\frac{1}{\sqrt{3}}J_z$. $J_i$
$(i=x,y,z)$ represent the spin-3/2 angular momentum matrices and cp
denotes the cyclic permutation of the preceding terms.

The Bir-Pikus strain Hamiltonian $H_{\epsilon}$ reads\cite{winkler2,bir} 
\begin{eqnarray}
H_{\epsilon}&=&{\rm diag}(-E_S^1,E_S^1,E_S^1,-E_S^1)-E_S^2I_4,\\
E_S^1&=&2D_u^{\prime}\epsilon_a,\label{eq11}\\
E_S^2&=&3D_d\epsilon_b,\label{eq12}\\
\epsilon_a&=&-\frac{1}{3}(1+\frac{1}{\sigma^{(111)}})\epsilon_{\|},\\
\epsilon_b&=&\frac{1}{3}(2-\frac{1}{\sigma^{(111)}})\epsilon_{\|},\\
\epsilon_{\|}&=&\frac{a_s-a_e}{a_e},\\
\sigma^{(111)}&=&\frac{C_{11}+2C_{12}+4C_{44}}{2C_{11}+4C_{12}-4C_{44}}.
\end{eqnarray}
Here, $D_u^{\prime}$ and $D_d$ stand for the deformation potential
constants; $\epsilon_b$ and $\epsilon_a$ denote the diagonal and off-diagonal components of
the strain tensor; $a_e$ and $a_s$ represent the lattice constants of epilayer (GaAs) 
and substrate materials (AlAs/InP); and $C_{11}$, $C_{12}$ and $C_{44}$ are the stiffness
constants.\cite{winkler2,bir}

\section{SPIN MIXING $\hat{\Lambda}_{{\bf k},{\bf k^{\prime}}}$
  FOR THE LOWEST HH AND LH SUBBANDS}\label{appC}
The spin mixing $\hat{\Lambda}_{{\bf k},{\bf k^{\prime}}}$ in the
spin-flip scattering induced by the EY mechanism reads $\hat{\Lambda}_{{\bf k},{\bf k^{\prime}}}=\hat{I}-\frac{1}{2}\sum_{n_z}(S_{\bf
  k}^{(1)}{S_{\bf k}^{(1)}}^{\dagger}-2S_{\bf k}^{(1)}{S_{\bf
    k^{\prime}}^{(1)}}^{\dagger}+S_{\bf k^{\prime}}^{(1)}{S_{\bf
    k^{\prime}}^{(1)}}^{\dagger})$, with $\hat{I}$ being a $2\times 2$
unit matrix. For the lowest hole subband, which is HH like, in GaAs/AlAs QWs, $S_{\bf k}^{(1)}$ is given by
\begin{eqnarray}
S_{\bf k}^{(1)}=\left(\begin{array}{cccc}
0 & -\frac{H_{14}^{1n_z}}{\Delta_{1n_z}^{hh}} &
-\frac{H_{12}^{1n_z}}{\Delta_{1n_z}^{hl}} &
-\frac{H_{13}^{1n_z}}{\Delta_{1n_z}^{hl}} \\
\\
-\frac{H_{41}^{1n_z}}{\Delta_{1n_z}^{hh}} & 0 &
-\frac{H_{42}^{1n_z}}{\Delta_{1n_z}^{hl}} &
-\frac{H_{43}^{1n_z}}{\Delta_{1n_z}^{hl}} 
\end{array}\right).
\end{eqnarray}
$H_{ij}^{1n_z}=\langle 1|H_{ij}|n_z\rangle$ in which $1$ ($n_z$)
stands for the first ($n_z$-th) subband and $H_{ij}$ represents the $i$-th row
and $j$-th column of the Hamiltonian matrix $H_{4\times 4}$ in Eq.~(\ref{eq6}).
$\Delta_{1n_z}^{hh}$ is the energy difference between the first and
$n_z$-th HH subbands and $\Delta_{1n_z}^{hl}$ is the energy difference
between the first HH and $n_z$-th LH subbands. 

For the lowest hole subband, which is LH like, in GaAs/InP QWs, $S_{\bf k}^{(1)}$ is given by
\begin{eqnarray}
S_{\bf k}^{(1)}=\left(\begin{array}{cccc}
0 & -\frac{H_{23}^{1n_z}}{\Delta_{1n_z}^{ll}} &
-\frac{H_{21}^{1n_z}}{\Delta_{1n_z}^{lh}} &
-\frac{H_{24}^{1n_z}}{\Delta_{1n_z}^{lh}} \\
\\
-\frac{H_{32}^{1n_z}}{\Delta_{1n_z}^{ll}} & 0 &
-\frac{H_{31}^{1n_z}}{\Delta_{1n_z}^{lh}} &
-\frac{H_{34}^{1n_z}}{\Delta_{1n_z}^{lh}} 
\end{array}\right),
\end{eqnarray} 
where $\Delta_{1n_z}^{ll}$ is the energy difference between the first and
$n_z$-th LH subbands. It is noted that $S_{\bf k^{\prime}}^{(1)}$ can be obtained by replacing ${\bf k}$ in
$S_{\bf k}^{(1)}$ with ${\bf k^{\prime}}$.

\section{EXACT SOLUTIONS OF Eqs.~(\ref{eq19}) TO (\ref{eq21})}\label{appD}
\begin{eqnarray}
S_z&=&S_x(0)\frac{\langle\Omega_x\Omega_z\rangle}{u}(-e^{-\lambda_1\tau_p^{*}t}+e^{-\lambda_2\tau_p^{*}t})+
\frac{S_z(0)}{2u}[(\langle\Omega_x^2\nonumber\\&&\mbox{}-\Omega_z^2\rangle+u)e^{-\lambda_1\tau_p^{*}t}+
(\langle\Omega_z^2-\Omega_x^2\rangle+u)e^{-\lambda_2\tau_p^{*}t}],\label{eq16}\nonumber\\\\
S_x&=&\frac{S_x(0)}{2u}[(\langle\Omega_z^2-\Omega_x^2\rangle+u)e^{-\lambda_1\tau_p^{*}t}+
(\langle\Omega_x^2-\Omega_z^2\rangle+u)\nonumber\\
&&\mbox{}\times e^{-\lambda_2\tau_p^{*}t}]+S_z(0)\frac{\langle\Omega_x\Omega_z\rangle}{u}
(-e^{-\lambda_1\tau_p^{*}t}+e^{-\lambda_2\tau_p^{*}t}),\label{eq17}\nonumber\\\\
S_y&=&S_y(0)e^{-\langle\Omega_x^2+\Omega_z^2\rangle\tau_p^{*}t},\label{eq18}
\end{eqnarray}
with
$\lambda_{1,2}=\frac{1}{2}(\langle\Omega_x^2+2\Omega_y^2+\Omega_z^2\rangle\pm
u)$ and
$u=\langle\Omega_x^2\rangle+\langle\Omega_z^2\rangle$.
\end{appendix}

\end{document}